\documentclass[11pt]{article}
\usepackage[margin=1in]{geometry}
\usepackage{amsmath,amssymb}
\usepackage{graphicx}
\usepackage{microtype}
\usepackage[hidelinks]{hyperref}
\graphicspath{{figures_acqvar/}}
\title{\vspace{-1.5em}\textbf{Acquisition state behaves as a structured, measurable
variable governing lung-nodule AI: kernel-driven measurement instability and
noise-driven detection fragility, invisible to DICOM metadata}}
\author{Daniel Soliman}
\date{}

\begin{document}
\maketitle

\begin{abstract}
\noindent\textbf{Background.} Governance for AI in imaging is rapidly formalizing:
the 2026 ACR--SIIM Practice Parameter for Imaging Artificial Intelligence
recommends local acceptance testing before deployment and ongoing real-world
performance monitoring with predefined stop rules \cite{acrsiim}. Current AI
quality programs---including the ACR Assess-AI registry---monitor \emph{outputs}
(AI-to-report concordance) using DICOM \emph{metadata} for context. We argue that
a necessary, currently-unmonitored layer sits \emph{underneath} output metrics:
whether incoming studies remain within the \emph{acquisition envelope} under which
a model was validated. We test whether acquisition state behaves as a structured,
measurable variable with distinct, axis-specific effects on AI behavior, and
whether those effects are recoverable from image metadata.

\noindent\textbf{Methods.} Using a MONAI RetinaNet lung-nodule detector
(LUNA16-trained), we evaluated (a) real paired CT acquisitions differing only in
reconstruction kernel (NLST B30f/soft vs.\ B80f/sharp; 155 matched nodules), and
(b) controlled multi-condition perturbations of LIDC-IDRI (dose, reconstruction
kernel, slice thickness) where one acquisition axis varies at a time. AI-reported
nodule diameter, detection confidence, and Fleischner size-category crossings were
measured. Acquisition state was characterized directly from pixels with a
4-feature noise/frequency fingerprint, whose information content we compared to the
ConvolutionKernel DICOM tag (with a QIBA CT phantom as a metadata-controlled
reference).

\noindent\textbf{Results.} On \emph{real} kernel pairs, reconstruction kernel alone
shifted AI-measured diameter (mean $|\Delta|$ 0.27\,mm) and flipped a Fleischner
size category in \textbf{5.2\% (8/155)} of nodules (same patient, same
acquisition), concentrated in 6--10\,mm nodules where size thresholds fall, while
detection confidence was statistically unchanged (Wilcoxon $p=0.22$). In controlled
perturbations, the effects dissociated by axis: the \textbf{noise axis} degraded
detection confidence (noise vs.\ frequency, $p=5.9\times10^{-32}$, concentrated in
$<$6\,mm nodules, $p=4.2\times10^{-25}$) but not measurement; the
\textbf{frequency/kernel axis} corrupted measurement ($p=8.6\times10^{-13}$) but
not detection. A pixel-derived acquisition fingerprint recovered reconstruction
identity (patient-level AUC $\approx$0.95 on real CT; 0.995 on the QIBA phantom)
where the ConvolutionKernel tag was uninformative (identical ``STANDARD'' labels
across reconstructions).

\noindent\textbf{Conclusion.} Acquisition state appears to be a structured variable
whose axes map to distinct AI failure modes---frequency content governs measurement
reliability, noise governs detection sensitivity---and these are not recoverable
from DICOM metadata alone. Output- and metadata-based monitoring can detect that
performance changed but cannot attribute it to acquisition drift. Acquisition-aware,
input-side validation is the missing layer for the acceptance-testing and
drift-monitoring requirements now entering imaging-AI accreditation.
\end{abstract}

\section{Introduction}
Imaging AI is moving from research into accredited clinical practice. The 2026
ACR--SIIM Practice Parameter for Imaging Artificial Intelligence---a
practice-parameter-class document---recommends that facilities perform local
acceptance testing prior to deployment, maintain an inventory of tools and
versions, monitor real-world performance for drift and safety, and define stop
rules. Facilities that adopt these practices can earn the voluntary ACR Recognized
Center for Healthcare-AI (ARCH-AI) designation, a precursor to a formal
accreditation anticipated in 2027 \cite{acrsiim}. In parallel, the ACR Assess-AI
registry operationalizes post-deployment monitoring by measuring concordance
between AI outputs and radiology-report-derived surrogate labels, alongside
anonymized patient demographics and exam metadata \cite{assessai}.

Although an important first step in AI oversight for clinical imaging, these
programs are, by construction, \emph{output-} and \emph{metadata-}centric: they
answer ``did the AI's output match the report?'' and contextualize with header
fields. But before satisfactorily answering \emph{why} an AI's performance changed,
one must know whether the imaging \emph{inputs} changed---whether incoming studies
still fall within the acquisition conditions under which the model was tested and
validated.

This validation layer is non-trivial. First, acquisition state is not fully
recoverable from DICOM metadata: reconstructions that behave very differently can
carry identical header descriptors (e.g., the ConvolutionKernel tag reads
``STANDARD'' for both filtered-back-projection and iterative reconstructions).
Second, the perturbation simulations commonly used to \emph{test} robustness do not
necessarily reproduce the effects of real reconstruction operations---they can
over- or under-state the effect depending on the quantity measured. A monitoring
program blind to the input distribution can therefore detect a performance change
without being able to attribute it, and a validation program built on
mis-calibrated simulations can mis-estimate fragility.

Medical imaging already treats many sources of variation as variables. Patient age,
sex, body habitus, and disease burden are understood as dimensions along which
model behavior is expected to shift, and validation is designed around them.
Acquisition factors are usually treated differently, as protocol settings or
nuisance variability to be standardized away. A kernel is chosen, a dose is set, a
slice thickness is selected. Yet reconstruction kernel, dose, and slice thickness
each move an image through a physical space and change the information available to
a reader or a model \cite{antun}, often in ways a header field does not record.

This motivates a more fundamental question. Are these acquisition factors best
understood as isolated protocol descriptors, or as coordinates of an underlying
acquisition state that a model must be validated against, like any other variable?
If the latter, the variation should show structure. Distinct acquisition axes
should produce distinct and reproducible effects on model behavior; those effects
should be measurable from the images themselves rather than inferred from metadata;
and the resulting coordinates should retain their meaning across scanners built by
different manufacturers. We test these three predictions directly: an axis-specific
dissociation of failure modes, recovery of acquisition state from pixels where
metadata is blind, and transport of the kernel axis across four vendors.

\section{Methods}
\textbf{Detector.} MONAI \cite{monai} RetinaNet lung-nodule detection model
(LUNA16-trained \cite{luna16}); frozen weights throughout. Per-scan outputs: candidate boxes (center, extent) and
confidence scores.

\textbf{Real kernel pairs (NLST \cite{nlst}).} Paired reconstructions of the same raw
acquisitions differing only in kernel: B30f (soft/STANDARD) vs.\ B80f (sharp/LUNG).
155 nodules detected and matched across both kernels. For each matched nodule we
recorded AI-reported diameter under each kernel ($\Delta$ diameter), per-kernel
confidence, centroid displacement, and whether the diameter crossed a Fleischner
size threshold (6/8\,mm). For size-stratified crossing rates we additionally
aggregated an expanded cohort of 110 patients (367 matched nodules, one timepoint
per patient), matching nodules across kernels by mutual nearest-neighbor within
5\,mm with a 2\,mm size-consistency gate to exclude cross-structure mismatches;
crossings were counted only when the diameter shift exceeded the 0.41\,mm
longitudinal test--retest noise floor, and stratum rates are reported with
patient-clustered bootstrap confidence intervals.

\textbf{Controlled perturbations (LIDC-IDRI \cite{lidc}).} Single-axis acquisition
perturbations of baseline scans via a physics-guided degradation engine
\cite{beyondbench}: dose reduction (noise), reconstruction kernel (frequency), and
slice thickness (resolution). Per (case, condition) we recorded patient-level
detection outcome, per-nodule detection-confidence change
($\mathrm{score}_{\mathrm{baseline}} - \mathrm{score}_{\mathrm{condition}}$), and
AI-diameter change vs.\ baseline. $n\approx180$ cases.

\textbf{Acquisition fingerprint.} Four pixel-level patch features (noise $\sigma$,
gradient sharpness, high/low-frequency ratio, neighbor correlation) computed on
tissue patches; used (a) as a per-(case,condition) acquisition coordinate, and (b)
to classify reconstruction identity. Compared against the ConvolutionKernel DICOM
tag. A QIBA CT phantom \cite{qiba} (FBP vs.\ ASIR, identical kernel tags) served as a
metadata-controlled reference.

\textbf{Cross-vendor transportability.} To test whether the kernel axis is shared
across manufacturers or scanner-specific, we assembled paired soft- and
sharp-kernel reconstructions from four vendors (GE, Philips, Siemens, Toshiba; 737
matched nodules) and extracted the detector's internal feature embeddings at five
levels (two backbone layers, three feature-pyramid levels). Because the pairs are
matched per nodule, anatomy cancels in the soft-to-sharp difference, isolating the
kernel effect. We assessed transportability two ways. First, shift direction: per
vendor we took the mean soft-to-sharp shift vector and measured its pairwise cosine
similarity across vendors. Second, leave-one-vendor-out (LOVO) transport: after
centering out each vendor's baseline, we trained a logistic kernel discriminator
(soft vs.\ sharp) on three vendors and tested its ROC-AUC on the held-out fourth. A
within-vendor five-fold cross-validated AUC, computed per vendor, served as the
ceiling.

\textbf{Analysis.} Measurement effect $= |\Delta\,\mathrm{diameter}|$; detection
effect $=$ confidence change and patient-level miss. Axis-specific effects compared
across condition families (noise/frequency/resolution) by nonparametric tests;
size-stratified. Real-kernel detection-confidence symmetry tested by Wilcoxon
signed-rank. Fingerprint vs.\ metadata compared by ROC-AUC for reconstruction
classification.

\section{Results}

\subsection{Real reconstruction-kernel change shifts AI measurement and flips
clinical categories, without affecting detection}
\begin{figure}[ht]
\centering
\includegraphics[width=0.62\textwidth]{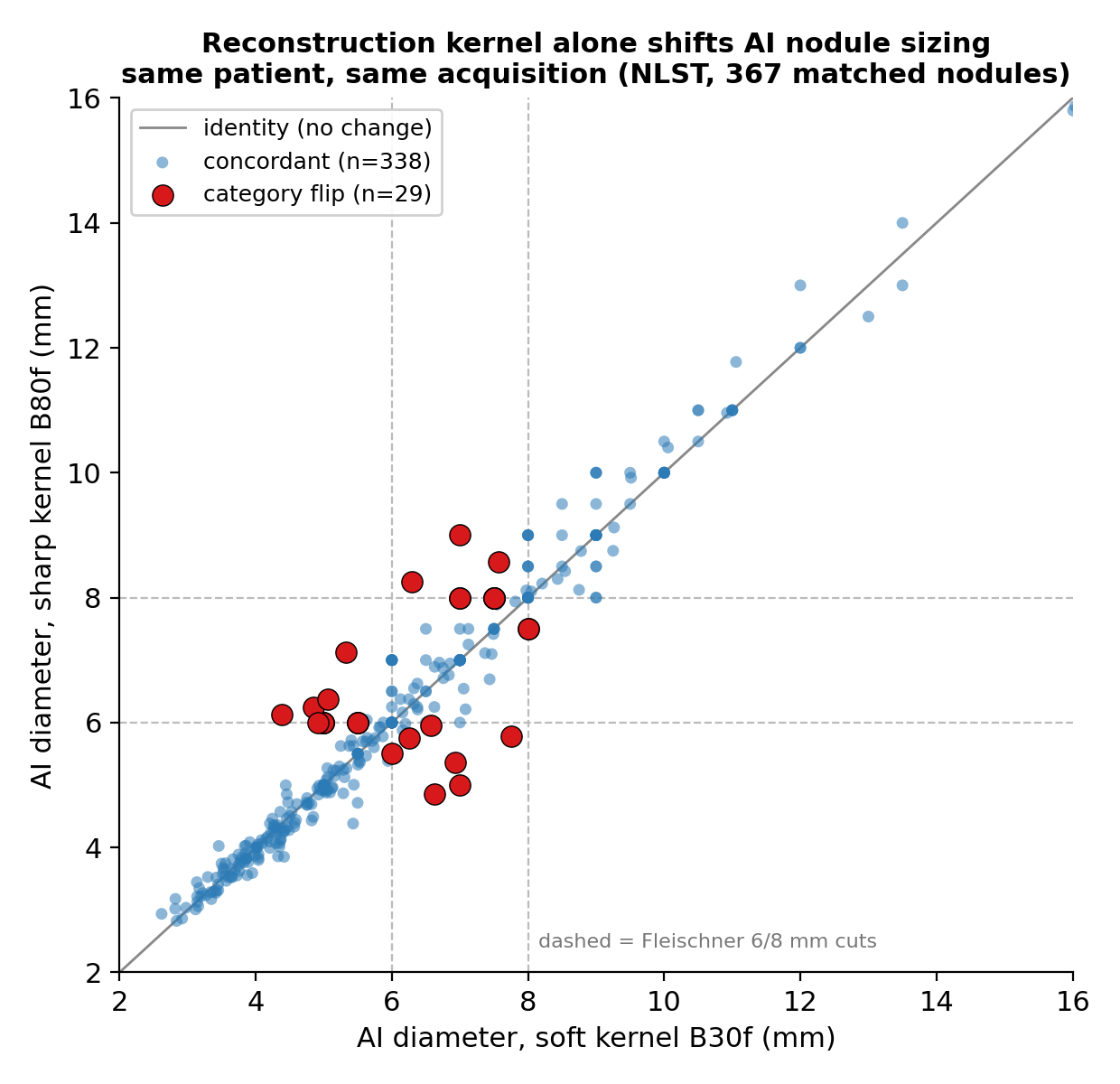}
\caption{Reconstruction kernel alone shifts AI-measured nodule diameter on real
paired NLST acquisitions (same patient, same raw scan, soft B30f vs.\ sharp B80f).
Each point is one matched nodule; red points cross a Fleischner 6 or 8\,mm
management threshold (diameter shift exceeding the 0.41\,mm test--retest noise
floor). Crossings concentrate at the 6/8\,mm cuts, where a sizing change alters the
follow-up recommendation.}
\label{fig:disc}
\end{figure}

Across 155 matched nodules on real B30f/B80f pairs, kernel change alone produced a
mean $|\Delta\,\mathrm{diameter}|$ of 0.27\,mm and crossed a Fleischner
size-category threshold in \textbf{5.2\% (8/155)} of nodules---same patient, same
acquisition, reconstruction kernel the only difference. The crossings concentrated
in the smaller nodules where management thresholds fall: in an expanded cohort (110
patients, 367 matched nodules), \textbf{12\% (95\% CI 6--19\%, patient-clustered) of
6--10\,mm nodules} crossed a category from kernel alone, and 5.9\% of 3--6\,mm
nodules, versus 0\% of nodules larger than 10\,mm (Figure~\ref{fig:disc}). Detection
confidence was statistically unchanged between kernels (Wilcoxon $p=0.22$): nodules
detected under one kernel were confidently detected under the other. Kernel change
is thus a \emph{measurement}-domain perturbation with direct clinical consequence
(altered follow-up recommendation), not a detection-domain one.

\subsection{Acquisition effects dissociate by physical axis}
\begin{figure}[ht]
\centering
\includegraphics[width=0.85\textwidth]{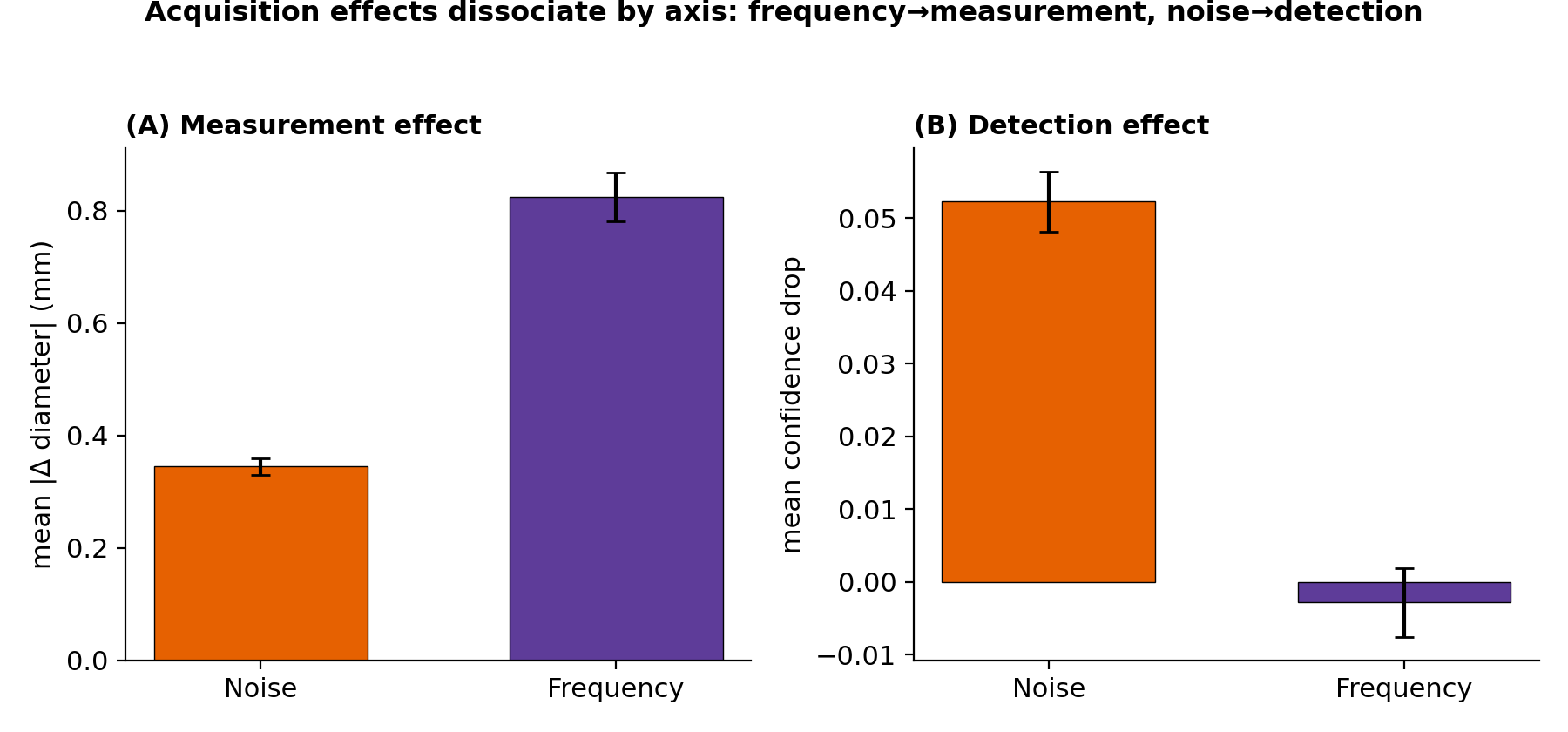}
\caption{Acquisition effects dissociate by physical axis. (A) Measurement effect
(mean $|\Delta\,\mathrm{diameter}|$) is driven by the frequency/kernel axis, not
the noise axis. (B) Detection effect (mean confidence drop) is driven by the noise
axis, not the frequency axis. Each axis is strong on its own failure mode and
near-null on the other. Error bars: SEM.}
\label{fig:axis}
\end{figure}

Under controlled single-axis perturbations, the failure modes separated by axis
(Figure~\ref{fig:axis}):
\begin{itemize}
\item \textbf{Noise axis (dose) $\to$ detection.} Detection confidence dropped
under noise but not under frequency change (noise vs.\ frequency,
$p=5.9\times10^{-32}$), with the effect concentrated in small ($<$6\,mm) nodules
($p=4.2\times10^{-25}$)---i.e., noise pushes small nodules toward the miss
threshold. Noise had no measurement effect.
\item \textbf{Frequency axis (kernel) $\to$ measurement.} Reconstruction kernel
drove diameter discordance (frequency vs.\ noise, $p=8.6\times10^{-13}$) but did
not shift detection confidence.
\end{itemize}
The dominant effects separated by axis: measurement effects concentrated on the
frequency axis, detection-confidence effects on the noise axis, each largely null
on the other. This indicates that acquisition state is not a single scalar
``quality'' but a structured vector whose components govern different downstream
behaviors. (We frame the detection arm as a confidence precursor rather than a
demonstrated miss rate; see Limitations.)

\subsection{The structure is recoverable from pixels but not from metadata}
\begin{figure}[ht]
\centering
\includegraphics[width=0.72\textwidth]{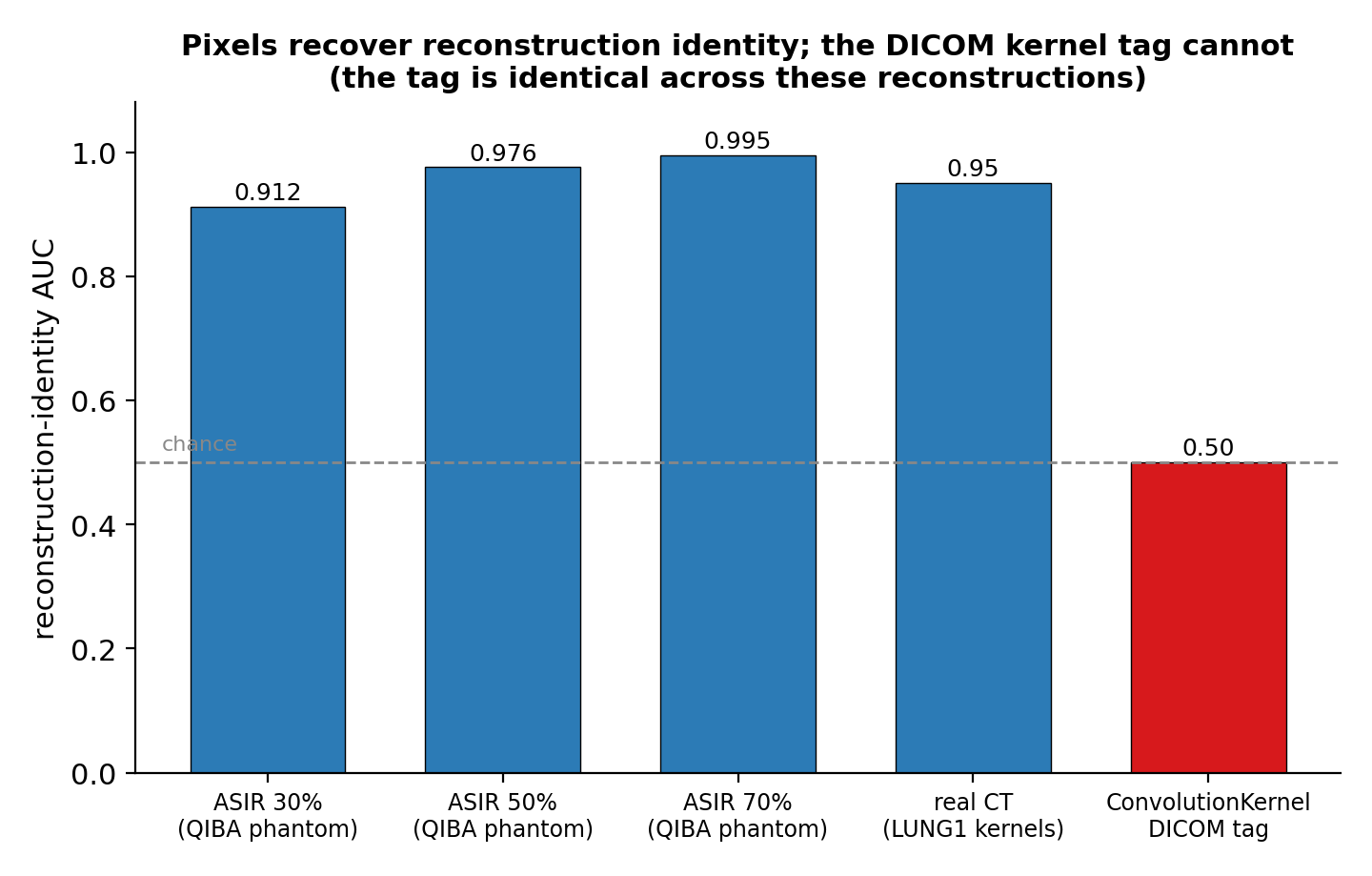}
\caption{Reconstruction identity is recoverable from pixels but not from the DICOM
header. A 4-feature pixel fingerprint separates reconstructions with rising AUC as
they diverge (QIBA FBP vs.\ ASIR 30/50/70\%: 0.91/0.98/0.99; real CT kernels
$\approx$0.95), whereas the ConvolutionKernel tag is identical across these
reconstructions and sits at chance.}
\label{fig:pixmeta}
\end{figure}

The pixel-derived acquisition fingerprint classified reconstruction identity at
patient-level AUC $\approx$0.95 on real CT and 0.995 on the QIBA phantom, while the
ConvolutionKernel DICOM tag was uninformative (identical labels across
reconstructions that the AI treats very differently). Metadata-based context
therefore cannot see the axis along which the AI is most affected; the input
distribution must be characterized from the images themselves (Figure~\ref{fig:pixmeta}).

\subsection{The kernel axis is shared across manufacturers}
\begin{figure}[ht]
\centering
\includegraphics[width=0.95\textwidth]{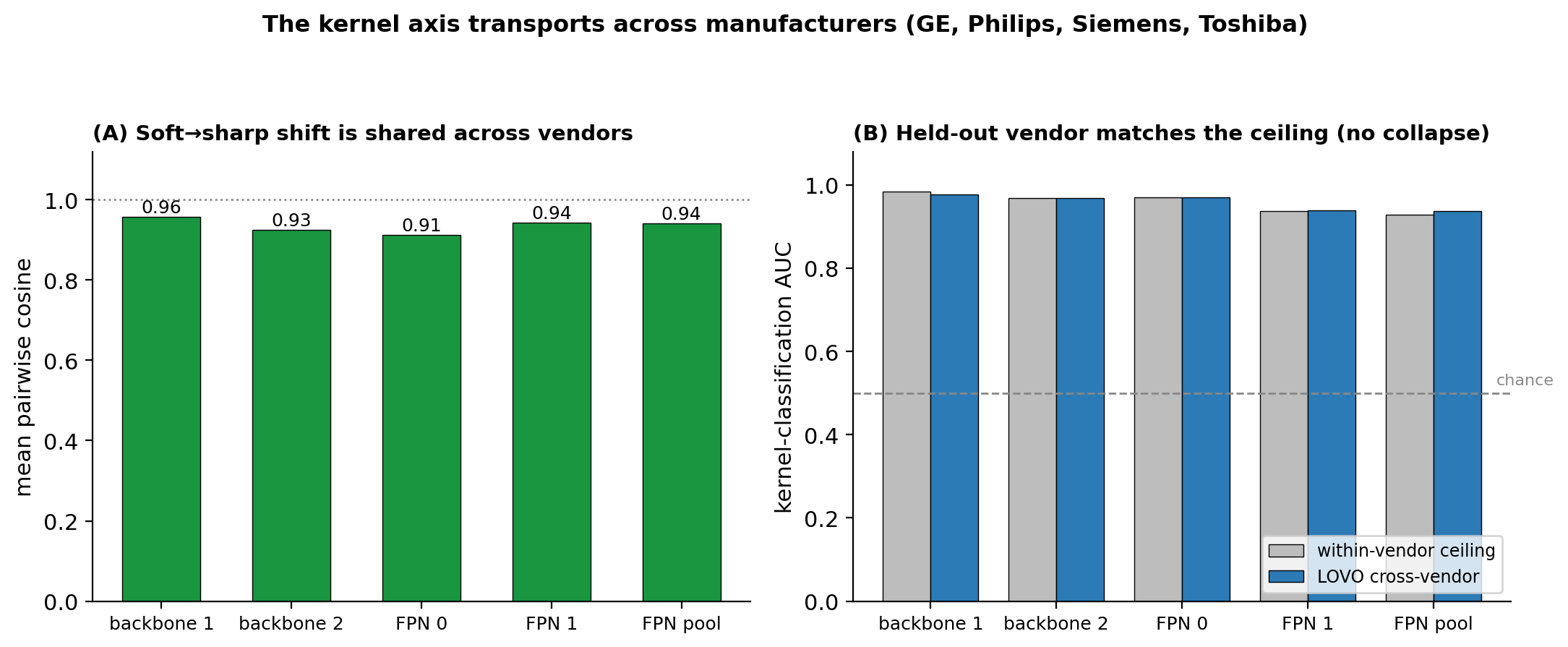}
\caption{The kernel axis transports across manufacturers. (A) The mean
soft$\to$sharp shift vector is near-parallel across GE, Philips, Siemens, and
Toshiba (pairwise cosine 0.91--0.96 by feature level). (B) A kernel discriminator
trained on three vendors and tested on the held-out fourth (leave-one-vendor-out)
matches the within-vendor ceiling at every level (AUC 0.94--0.98 vs.\ 0.93--0.98),
with no out-of-vendor collapse.}
\label{fig:vendor}
\end{figure}

The kernel axis transports across vendors rather than being scanner-specific
(Figure~\ref{fig:vendor}). The
soft-to-sharp shift was near-parallel across the four manufacturers, with pairwise
cosine similarity of 0.91 to 0.96 by feature level, indicating a common translation
rather than per-vendor drift. In leave-one-vendor-out testing, a kernel
discriminator trained on three vendors classified the held-out fourth at AUC 0.94
to 0.98, statistically indistinguishable from the within-vendor ceiling of 0.93 to
0.98. Per held-out vendor, AUC ranged from 0.90 (Toshiba) to 1.00 (Siemens), with
no out-of-vendor collapse at any feature level. A model that never saw a given
manufacturer still separated that manufacturer's soft and sharp reconstructions at
near-ceiling accuracy. The discriminating coordinate is therefore shared across
scanners engineered independently, consistent with acquisition state being a common
variable rather than a per-scanner diagnostic.

\section{Discussion}
Acquisition variation appears to be structured rather than random. Its
frequency-content axis governs \emph{measurement} reliability and, because nodule
size determines Fleischner category, a kernel change can flip a follow-up
recommendation with no change in the patient and no detection failure to flag it.
Its noise axis governs \emph{detection} sensitivity, most acutely for small
nodules. The two axes are separable and act through different mechanisms.

A further observation suggests that acquisition state is a genuine \emph{variable}
rather than an individual scanner idiosyncrasy: its structure appears to be
invariant across manufacturers. In a leave-one-vendor-out analysis spanning four
vendors, the soft$\to$sharp shift that reconstruction kernel induces in the
detector's internal representation was near-parallel across vendors (pairwise
cosine 0.91--0.96 across feature levels) and a kernel discriminator trained on three
vendors transported to the held-out fourth with essentially no loss (cross-vendor
AUC 0.94--0.98, matching the within-vendor ceiling). We had expected kernel effects
to be strongly vendor-specific; instead the signal transported across manufacturers
with minimal degradation. The kernel axis is thus not a collection of
scanner-specific quirks but a shared, transportable coordinate. A model that has
never encountered a given manufacturer still locates that manufacturer's
reconstructions along the same axis. These findings are consistent with acquisition
state behaving as a shared coordinate rather than a collection of vendor-specific
effects. The practical corollary for validation is that an acquisition envelope
characterized on one fleet need not be re-derived from scratch on another. We frame
this as preliminary: the result is established here for the network's internal
embedding, and confirming it for the pixel-level fingerprint that a deployed,
model-agnostic monitor would use remains future work.

This has direct implications for the governance layer now being standardized.
Output-concordance monitoring (Assess-AI) and metadata context can establish
\emph{that} performance drifted; they cannot, on their own, distinguish
acquisition-driven drift from case-mix drift, because the discriminating signal
lives in the pixels, not the report or the header. The Practice Parameter's
recommendations---local acceptance testing and drift monitoring with stop
rules---implicitly require an \emph{input-side} check: is this study within the
acquisition envelope the model was validated under? Acquisition-aware validation
supplies that layer, and the dissociation shows it must be at least two-dimensional
(a measurement-reliability axis and a detection-reliability axis), reported per
failure mode rather than as a single score.

\section{Limitations}
\begin{itemize}
\item \textbf{Single detector} (one LUNA16-trained RetinaNet); generalization to
other architectures/tasks is untested. Effects are consistent across both real and
simulated stimuli for this model.
\item \textbf{Detection arm is a confidence precursor.} At the perturbation
magnitudes studied, the noise-axis effect manifests as reduced detection
\emph{confidence} (and increased proximity to the miss threshold), not as frequent
outright misses, which remained rare on predominantly solid nodules. We frame it as
an early-warning signal, not a demonstrated miss rate.
\item \textbf{Simulated magnitudes are uncalibrated to real MTF, and the mismatch
is feature-dependent.} The controlled perturbations isolate the \emph{structure} of
acquisition effects; their magnitudes are not calibrated to any specific scanner.
The simulated kernel produced a \emph{larger} diameter shift ($\sim$1\,mm) than the
real B30f/B80f pair (0.27\,mm), yet moved the high-frequency noise texture
\emph{less} than reality: the real soft-to-sharp change displaced the pixel
frequency-content feature roughly $8\times$ more than the simulated pair, and raised
high-frequency noise-power-spectrum amplitude $\sim$$5\times$. The direction of the
simulation's error therefore depends on the quantity measured---overstating the
geometric (diameter) effect while understating the textural (frequency) one---so
only the structure, not the magnitude, transfers between simulated and real
reconstruction.
\item \textbf{Direction is bidirectional.} Acquisition change produces
\emph{discordant} recommendations in both directions (over- and under-call), not
one-directional ``misses''; consequences should be described as recommendation
discordance.
\item \textbf{Datasets.} LIDC-IDRI and NLST; predominantly solid nodules;
subsolid/GGO behavior (expected to be more fragile) under-sampled.
\end{itemize}

\section*{Author Contributions}
D.S. conceived the study, developed the pixel-based acquisition fingerprint and the
MONAI detection experiments, performed the analyses, and drafted the manuscript.
Vera D.P. selected the QIBA reference dataset used for the metadata-controlled
comparison, contributed the framing distinguishing reconstruction-kernel
confirmation from real-world generalization, and advised on dataset suitability and
licensing. Both authors reviewed and approved the final manuscript.

\section*{Acknowledgments}
The authors thank the Quantitative Imaging Biomarkers Alliance (QIBA) and The Cancer
Imaging Archive (TCIA) for the phantom dataset, the National Cancer Institute for
access to NLST data, and the LIDC-IDRI investigators for the public lung-nodule
dataset.

\end{document}